\documentclass[aps,pra,twocolumn]{revtex4-1}

\usepackage{epstopdf}
\usepackage{subfigure}
\usepackage{dsfont}
\usepackage{graphicx}
\usepackage{color}

\begin{document}

\author{Micha\l{} Jachura}
\email{michal.jachura@fuw.edu.pl}

\author{Jan Szczepanek}

\author{Wojciech Wasilewski}

\author{Micha\l{} Karpi\'{n}ski}

\affiliation{Faculty of Physics, University of Warsaw, Pasteura 5, 02-093 Warszawa, Poland}

\title{Measurement of radio-frequency temporal phase modulation using spectral interferometry}


\begin{abstract}
We present an optical method to measure radio-frequency electro-optic phase modulation profiles by employing spectrum-to-time mapping realized by highly chirped optical pulses. We directly characterize temporal phase modulation profiles of up to $12.5$~GHz bandwidth, with temporal resolution comparable to high-end electronic oscilloscopes. The presented optical setup is a valuable tool for direct characterization of complex temporal electro-optic phase modulation profiles, which is indispensable for practical realization of deterministic spectral-temporal reshaping of quantum light pulses.
\end{abstract}

\maketitle

\section{Introduction}
Encoding of quantum information in the spectral-temporal degree of freedom of optical pulses has been recently recognized as a promising platform for high-dimensional quantum information processing and quantum communication, that is naturally compatible with fibre-optic infrastructure and integrated optical setups \cite{Humphreys2013, Brecht2015, Lukens2017}. Realization of general spectral-temporal photonic quantum information processing and measurement requires spectral-temporal shaping of optical pulses that does not involve spectral or temporal filtering.
A recipe for such manipulations is well-known within the field of classical ultrashort pulse shaping \cite{Edition, Torres-Company2011,Salem}. It relies on successive imprints of spectral and temporal phases onto the optical pulse in a
controllable manner. While the former is achieved simply by propagating
the pulse through a dispersive medium, the latter requires elaborate
active modulation techniques, which may be implemented e.~g.\ by nonlinear-optical
\cite{Salem, Agha2014, Zhu13, Donohue2016, Matsuda2016} or electro-optic methods \cite{Poberezhskiy2005, Li2013, Jachura2016, Wright2017}. The substantial
advantage of electro-optic, and the recently demonstrated electro-optomechanical
approach \cite{Fan2016}, lies in its low-noise and deterministic operation, which are key characteristics for applications in optical quantum technologies \cite{Wright2017,Fan2016,Olislager,Weiner}.

In the electro-optic approach a well-defined temporal phase imprint is applied to the
optical pulse by driving an electro-optic phase
modulator (EOPM) with an appropriate radio-frequency (RF) field, of
up to tens of GHz bandwidths. Detailed characterization of
the imprinted temporal phase becomes thus a key necessity, especially given
recent efforts to diversify from standard single-tone RF modulation
techniques \cite{Olislager,Lansinis2015,Jachura2016,Fan2016} to more complex temporal phase profiles  \cite{Li2013,Poberezhskiy2005,Agha2016}, in particular in guided-wave electronic systems. Here we experimentally show the reconstruction of a fast temporal phase modulation profile performed using the spectral encoding method. The spectral encoding technique relies on mapping a temporal waveform of interest onto the spectrum of an optical pulse. This can be achieved by highly chirping the pulse and subsequently modulating either its spectral amplitude or spectral phase. The former variation is known as the photonic time stretch (PTS) technique and has been extensively used for the analysis of RF signals \cite{Jalali1999,Jalali2007,Jalali2013} as well as for detailed characterization of THz frequency electric-field pulses \cite{Jiang2011,THZ2000,THZ2011,THZ2015}. Since we are directly interested in the temporal phase profiles we use the latter variation of this method.

We use a highly chirped optical pulse to probe the phase imprinted by an RF field applied to the EOPM.  
The spectral phase of the optical pulse is afterwards recovered using standard spectral interferometry \cite{Takeda1982, Drescher} and, after appropriate chirp calibration,
translated into the temporal profile of the modulating pulse with
temporal resolution comparable to or surpassing the fastest electronic
oscilloscopes available. The temporal phase modulation profiles are directly probed without the need for electro-optic response calibration of the EOPM. Its calibration can be optionally employed
to translate the phase modulation profile into the electronic waveform
inducing it, which effectively turns the setup into a high-bandwidth
optical oscilloscope. 
 
Compared with the hitherto experiments using the spectral encoding approach our work is related to the
scheme presented in ref.\ \cite{Jiang2011}, where free-space THz waveforms were characterized. We extend the applications of THz-pulse-characterization methods to direct measurement of radio-frequency temporal phase modulation profiles in a guided-wave platform. Since our solution can also be employed for the reconstruction of RF signals themselves, it can be alternatively treated as a variation of PTS technique relying on phase rather than intensity modulation of a chirped optical pulse.

\section{Measurement scheme}

Our experimental implementation relies on a Mach-Zehnder interferometer
with a lithium-niobate-waveguide EOPM placed in
one of its arms and a variable delay line $\tau$ placed in the other
arm. The interferometer is fed with heavily chirped optical pulses and one of
its output ports is monitored by a spectrometer, such as an optical spectrum analyser (OSA). Assuming
the temporal mismatch between interferometer arms $\tau$, equal transmission of the interferometer arms over the wavelength range used in the experiment, the spectral phase imprint resulting from temporal RF modulation $\varphi(\omega)$ through time-to-frequency mapping induced by the pulse chirp, the spectral intensity of the output field is given by:
\begin{eqnarray*}
I(\omega) & \propto & |E(\omega)+E(\omega)\mathrm{e}^{i\omega\tau+i\Delta\psi(\omega)+i\varphi(\omega)}|^{2}=\\
 & = & 2|E(\omega)|^{2}+2|E(\omega)|^{2}\cos\left[\omega\tau+\Delta\psi(\omega)+\varphi(\omega)\right],
\end{eqnarray*}
where $E(\omega)$ is the initial spectral amplitude of the pulse
entering the interferometer and $\Delta\psi(\omega)$ is the difference in spectral phases imprinted by the interferometer arms. In the Fourier domain, the two terms in
the above equation correspond to three distinct peaks, i.~e.\ the baseband
peak and two symmetrically located sidebands. By choosing one of the
sidebands, performing an inverse Fourier transform 
one can recover the spectral phase profile $\omega\tau+ \Delta\psi(\omega) + \varphi(\omega)$
\cite{Takeda1982,Drescher}, which is a standard phase retrieval method
used e.~g.\ in spectral interferometry for femtosecond pulse characterization
\cite{Iaconis1999}. Retrieval of $\varphi(\omega)$ requires calibrating the value of $\tau$ and the differential $\Delta\psi(\omega)$  term. The sum of both contributions can be easily determined by retrieving a reference spectral phase profile with the RF modulation switched off. 
The reference spectral phase profile is afterwards subtracted from the actual measurement with the RF signal switched on.
Additionally, this procedure avoids the need for precise spectral calibration of the OSA \cite{Dorrer1}. Once the reference phase profile for a given delay $\tau$ is known, the RF waveforms can be recovered from single spectral measurements. 

Provided the pulses entering the interferometer
are chirped such that temporal far-field condition is satisfied \cite{Torres-Company2011},
the spectral phase $\varphi(\omega)$ can be directly translated into
the temporal profile of RF modulation using the linear relation
$\omega=\omega_{0}+at$, where $a$ is the chirping rate and $\omega_{0}$
is the central angular frequency of the pulse. Both the parameters
can be easily determined experimentally, as we detail further in the
text.

\section{Experimental setup}

\begin{figure}[t!] 
\begin{center}
\includegraphics[scale=0.3]{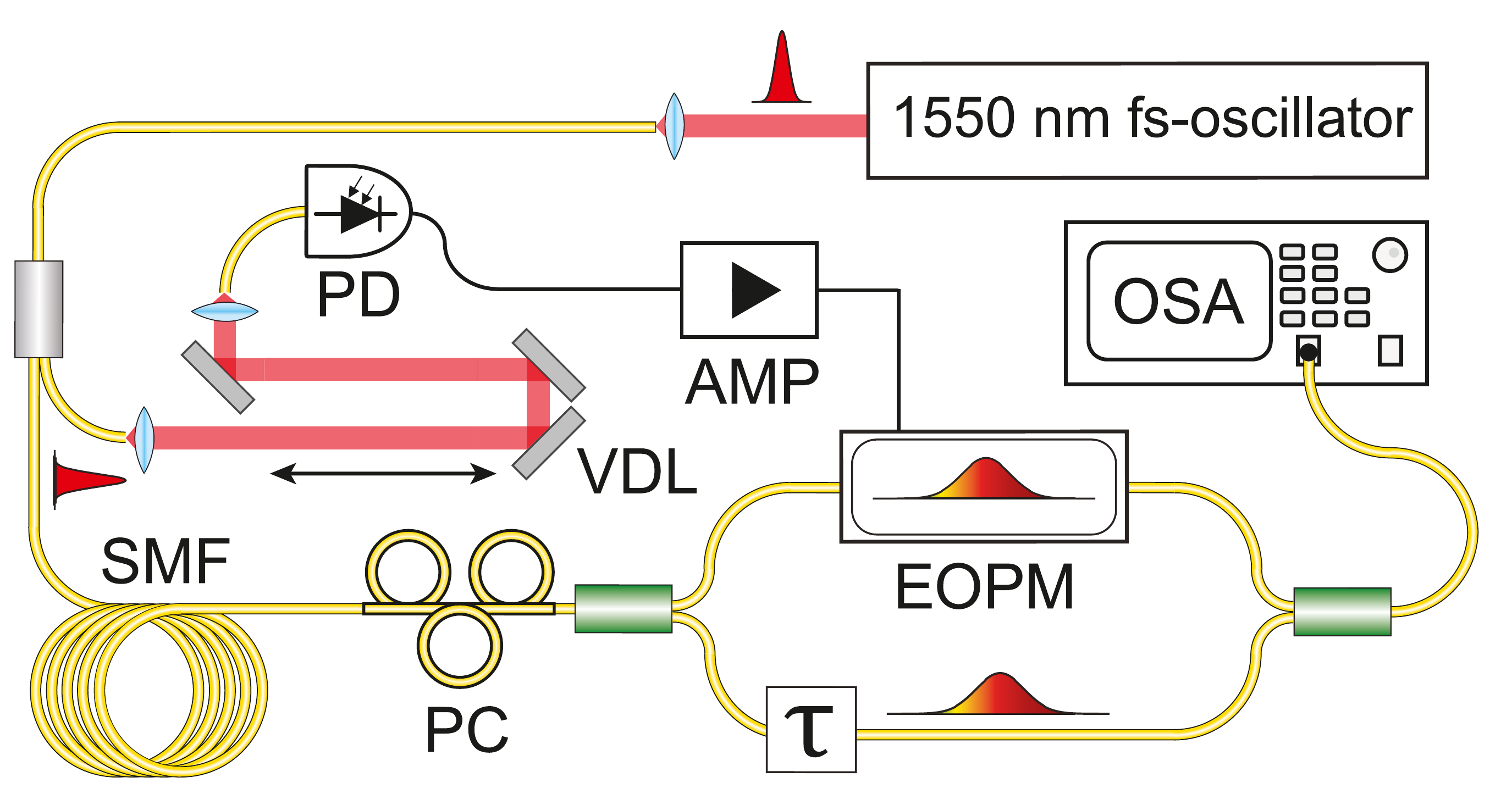}
\caption{Schematic representation of the experimental setup. PD, photodiode, AMP, amplifier, EOPM, electro-optic phase modulator, SMF, single mode fibre, PC, polarization controller, OSA, optical spectrum analyser, VDL; $\tau$, variable delay lines.} \label{Fig1}
\end{center}
\end{figure} 

The scheme of our experimental setup is presented in Fig.~\ref{Fig1}. We use
a polarization-maintaining-fibre Mach-Zehnder (MZ) interferometer with
the EOPM placed in one of its arms. Its advantage with respect to the intrinsically stable common-path polarization
interferometer \cite{Jiang2011,Kim2016} lies in the fact that optical field in the reference arm is not subjected to any temporal phase modulation. Additionally, in this setting the delay $\tau$ can be easily adjusted using a regular air-gap delay line. A pulsed beam from a fibre femtosecond oscillator (Menlo Systems C-Fiber HP, $1550$~nm central wavelength,
$50$~fs pulse duration) is fibre-coupled and divided in two parts, of
which one is sent through a free-space variable delay line and coupled into a $12.5$-GHz-bandwidth photodiode detector (PD, EoTech-3500FEXT),
generating the RF pulse to be characterized, while the other is chirped
in a length of single-mode fibre (SMF) and directed into the MZ interferometer.
The RF pulse generated by the PD is amplified by a pair of broadband
($0-40\,\mathrm{GHz}$) RF amplifiers connected in series and fed
into the EOPM (EO-Space PM-DV5-40-PFU-PFU-LV-UL)
placed in one of the interferometer arms. Inside the EOPM the
RF pulse temporally overlaps with the optical pulse chirped during
the propagation through either $250\,\mathrm{m}$ or $500\,\mathrm{m}$
of SMF (Corning SMF-28). The delay between the RF and
optical pulses is tuned using a $1.5\mathrm{\:m}$-long air-gap variable
delay line placed before the PD.

\begin{figure}[!b] 
\begin{center}
\includegraphics[scale=0.6]{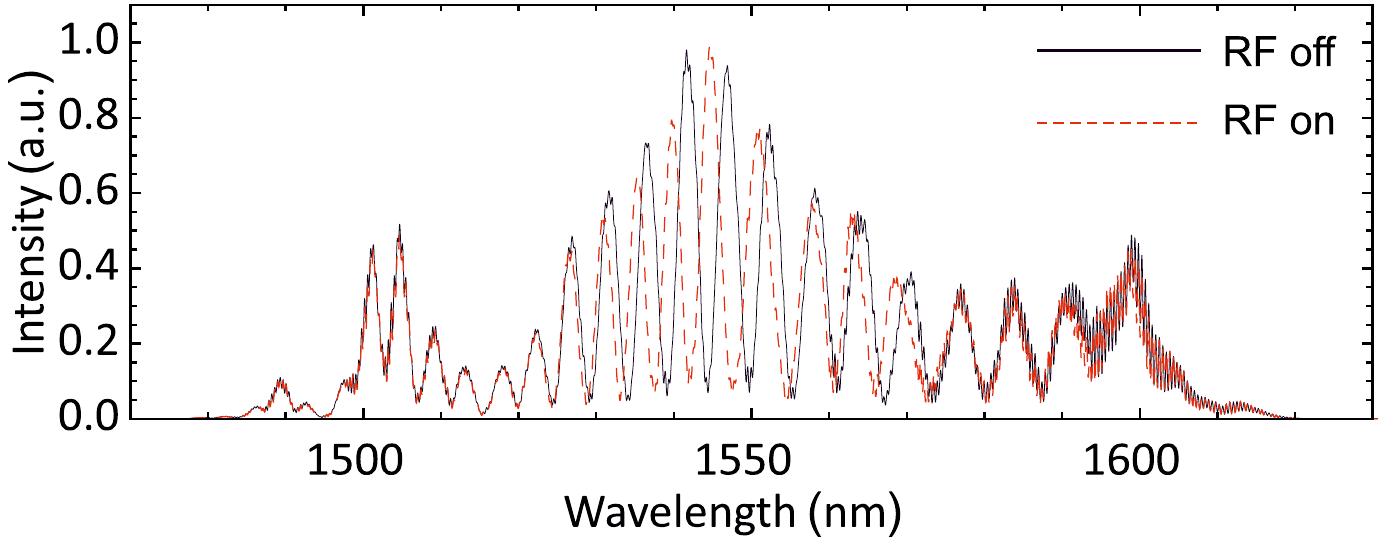}
\caption{Exemplary interferometer output spectra measured by the
 OSA for $\tau \neq 0$, with and without RF phase modulation.} \label{Fig2}
\end{center}
\end{figure} 

One of the output ports of the interferometer is monitored by an OSA (Yokogawa AQ6370D) whereas the second
output port is sent through a bandpass filter ($3$~nm full-width-half-maximum bandwidth) and directed
to an auxiliary photodiode detector (not presented in the experimental setup
scheme). Simultaneous monitoring of interference fringes in the spectral
and temporal domains simplifies the initial alignment of the interferometer
in a balanced position, which facilities the synchronization
between the RF and optical pulses. For the actual measurements the
interferometer is used in an unbalanced setting ($\tau \neq 0$), with only the OSA being monitored. 

The delay $\tau$, which defines the spectral fringe spacing, needs to be large enough to yield a clearly distinguishable sideband in Fourier domain and small enough to avoid detrimental effects resulting from the finite resolution of the spectrometer. Our delay setting resulted in fringe spacing of $1.5$~nm. Setting the resolution of the OSA to $0.1$~nm allowed us minimize the effects finite spectrometer resolution of data interpolation on the reconstructed spectral phase \cite{Dorrer2}.

\section{Results and discussion}

In Fig.~\ref{Fig2} we present exemplary interferometer output spectra $(\tau\neq0)$
collected using the OSA with RF modulation switched off and on for
the case of a $250$-m-long chirping SMF. The phase
imprinted by the RF pulse significantly modifies the positions of
minima and maxima of spectral interference fringes over bandwidth range of approximately
$50\mathrm{\,nm}$. In further experiments we used
larger values of $\tau,$ resulting in denser spectral fringes,
which allowed us to easily separate one of the spectral sidebands
after its transformation to the Fourier domain. In our experiment
the acquisition time of a single OSA measurement is much longer than
the time interval between two subsequent optical pulses. Single-shot acquisition could be achieved by employing a spectrometer with a multipixel detector array and reducing the repetition rate of the fs oscillator to match the spectrometer 
refresh rate (e.~g.\ by an external pulse picker) \cite{Kim2016,Drescher}. Alternatively, in analogy to the PTS approach, frequency-to-time mapping via the dispersive Fourier transform principle could be used to record the spectrum using a single fast photodiode \cite{Jalali1999}, possibly with an appropriate reduction of repetition rate.

\begin{figure}[tb] 
\begin{center}

\includegraphics[scale=0.5]{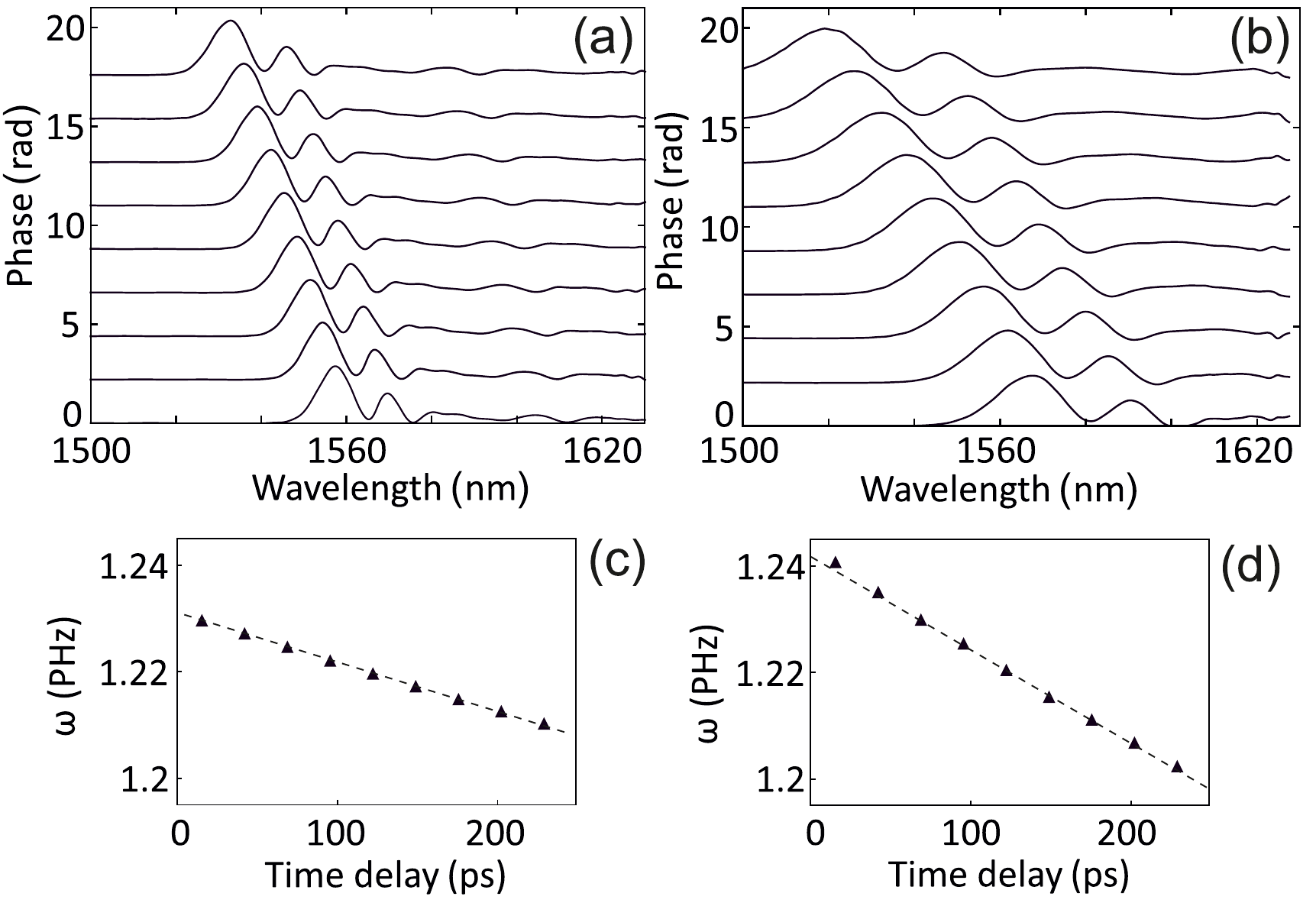}
\caption{Phase modulation profiles retrieved for a series of different delays between the RF waveform and the optical pulse. The maxima of the waveforms are traced to recover the SMF group delay dispersion value
yielding $-11.1\,\mathrm{ps\,^{2}}$ for $500\,\mathrm{m}$ of SMF~(a,~c) and $-5.73\,\mathrm{ps\,^{2}}$ for $250\,\mathrm{m}$
of SMF~(b,~d).} \label{Fig3}

\end{center}
\end{figure}

\begin{figure}[ht] 
\begin{center}

\includegraphics[scale=0.6]{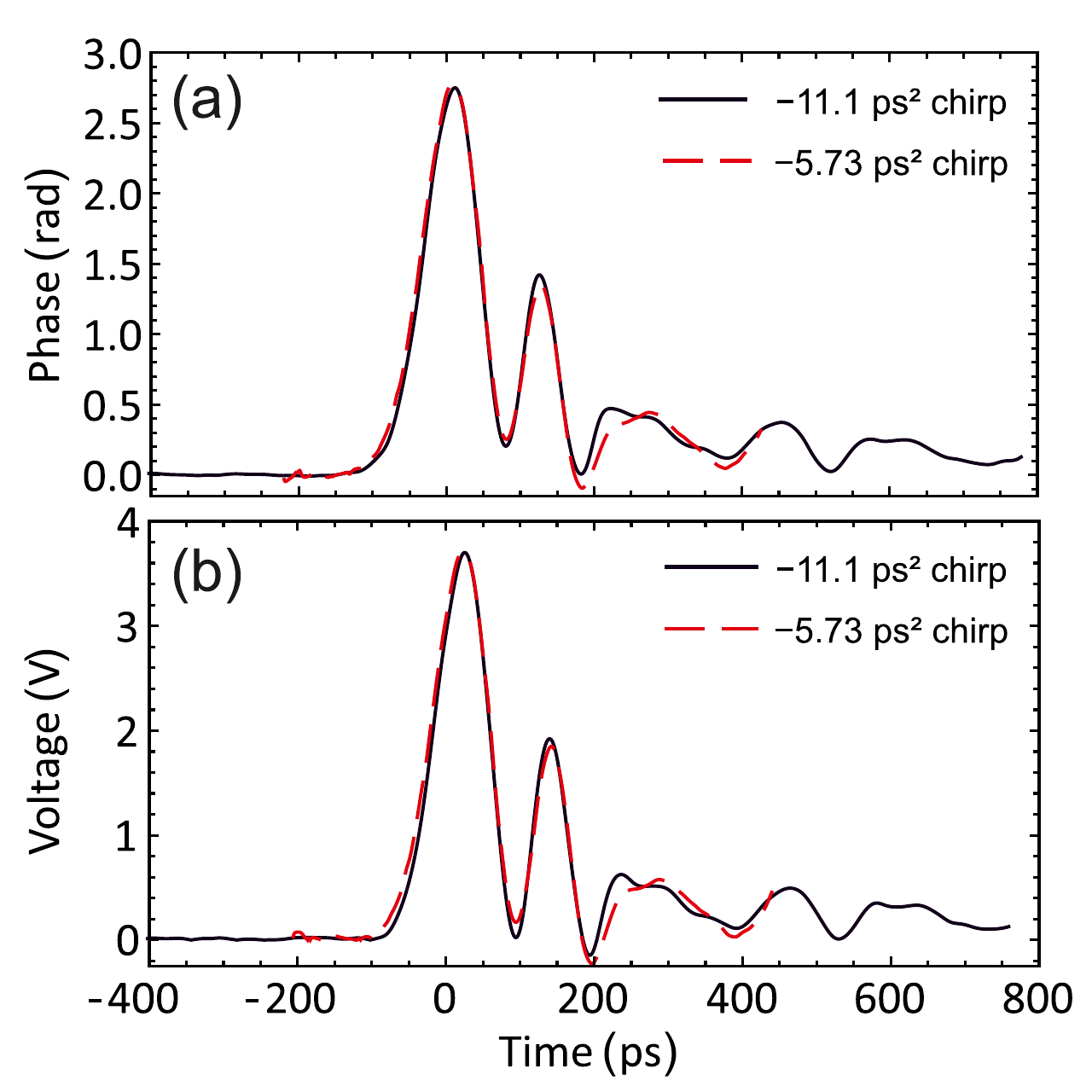}
\caption{Retrieved RF temporal phase modulation profile (a), 
and the corresponding electronic waveform profile (b) for the two values of
optical pulse chirp (solid and dashed lines).} \label{Fig4}

\end{center}
\end{figure} 

To retrieve the chirp value $a$ and verify the repeatability
of our measurement scheme we retrieved the phase modulation profiles
for a series of delays between the RF waveform and the
optical pulse. This has been achieved by adjusting the air-gap variable
delay line in a series of $4-\mathrm{mm}$ steps corresponding to
$13.34\,\mathrm{ps}$ of delay. In Fig.~\ref{Fig3}(a,b) we present the result
of this procedure for the $250\,\mathrm{m}$ and $500\,\mathrm{m}$
lengths of the chirping SMF. The position of modulation maximum
has been recorded to recover the SMF group delay dispersion values yielding 
$-11.1\,\mathrm{ps^{2}}$ for $500\,\mathrm{m}$ of SMF, Fig.~\ref{Fig3}(c),
and $-5.73\,\mathrm{ps^{2}}$ for $250\,\mathrm{m}$ of SMF, Fig.~\ref{Fig3}(d). 

The final result of the experiment is shown in Fig.~\ref{Fig4}, where we present
the RF modulation profiles retrieved for the two values of optical
pulse chirp, Fig.~\ref{Fig4}(a). The reconstructed waveforms agree very well with each other.
Naturally the temporal interval covered by the chirped optical pulse
is longer for a larger GDD value. We verified that the sensitivity of
our setup allows measuring pulses generated directly by the PD with a peak voltage of $100\,\mathrm{mV}$. 

For the purpose of spectral-temporal light shaping the most relevant
information is the temporal phase modulation profile induced by the
RF pulse. It can however be converted to retrieve the actual electronic
waveform by using the EOPM electro-optic response profile. We used the response profile provided the EOPM manufacturer in the $0-30\,\mathrm{GHz}$ frequency range. The EOPM response can be independently verified using spectral sideband analysis \cite{kalibracja}. In the temporal domain the measured phase modulation profile is a convolution of the electronic signal feeding the EOPM with its electro-optic response
function. Thus, the retrieval of electronic waveform requires a reciprocal
deconvolution operation, which is especially easy to implement in the
spectral domain, where it corresponds to division by the EOPM spectral
response function. The division performed in the spectral domain is also beneficial for the deconvolution fidelity since this is the domain where the instrument response function is directly measured. The results of the deconvolution procedure, based on the spectral response profile provided by the EOPM manufacturer, are presented in Fig.~\ref{Fig4}(b).

The temporal resolution achievable through chirped-pulse spectral-temporal mapping is given by $(t_\mathrm{b}t_\mathrm{a})^{1/2}$,
where $t_\mathrm{b}$ and $t_\mathrm{a}$ are the duration of the optical pulse before
and after the chirp \cite{Sun2016}. For the optical pulses employed in our setup this value equals approximately $8.7\,\mathrm{ps}$ for $500\,\mathrm{m}$ and $6.1\,\mathrm{ps}$ for $250\,\mathrm{m}$ of SMF. Although the expected optical temporal resolution can easily reach single ps, in practice the resolution is limited by the finite bandwidth of EOPMs. Whereas this limitation is irrelevant for the direct characterization of phase-modulation profiles, for probing of electronic waveforms it imposes a trade-off between the sensitivity and temporal resolution of the method, as the EOPM bandwidth could be increased at the cost of reduced sensitivity by using a shorter  modulator. This can be mitigated by the development of novel types of electro-optic modulators offering both exceptionally low half-wave voltage and high modulation bandwidth \cite{Enami,Melikyan}.    

\section{Conclusions and outlook}

We present a robust method for the direct characterization of RF electro-optic temporal phase modulation profiles by means of chirped-pulse sampling method. It extends the applications of THz field characterization methods and spectral encoding techniques to radio-frequency temporal phase profiles in an integrated optical platform, which are routinely employed for electro-optic spectral shaping of classical and quantum light. Alternatively it can be used for the reconstruction of RF electronic signals inducing phase modulation, which presents a variation of the PTS technique relying on phase rather than intensity modulation, involving a phase-sensitive detection scheme. Our work paves the way towards deterministic realization of complex spectral-temporal mode transformations, including general unitary transformations \cite{Morizur2010}, which will lead to important developments in spectrally-encoded photonic quantum information processing and communication \cite{Brecht2015}. Thanks to the reconfigurability of electro-optic phase modulation patterns we also expect it to become a valuable tool for realization of adaptive spectral-temporal shaping strategies, in particular of quantum light \cite{Bellini2012}.

\bigskip

\section*{Acknowledgements}

We thank K.~Banaszek, C.~Radzewicz and A. O. C. Davis for careful reading of the manuscript. This work has been supported by the National Science Centre of Poland (project no.\ 2014/15/D/ST2/02385). M.~J. was supported by the Foundation for Polish Science. The authors declare that they have no conflict of interest.

\end{document}